\documentclass[runningheads]{llncs}

\usepackage[T1]{fontenc}
\usepackage{amsmath,amssymb,amsfonts}
\usepackage{graphicx}
\usepackage{enumitem}
\usepackage{xcolor,colortbl,siunitx,pifont,subcaption}
\usepackage{booktabs}
\usepackage{multirow}
\usepackage{multicol}
\usepackage{array}
\usepackage{tabularx}
\usepackage{float}
\usepackage{hyperref}
\usepackage{xurl}  
\usepackage{tikz}
\usepackage[capitalise,nameinlink]{cleveref}

\definecolor{bestrow}{HTML}{EAF4EA}
\definecolor{hdrrow} {HTML}{EBF3FB}
\definecolor{hardcol}{HTML}{FFF3CD}

\newcommand{\best}[1]{\textbf{#1}}
\newcommand{\swatch}[2]{\colorbox[HTML]{#1}{\phantom{x}}\,\scriptsize#2}

\newcolumntype{V}{!{\vrule width 0.4pt}}

\begin{document}

\title{Learning Class Difficulty via Dynamic Focal Attention for Histopathology Segmentation}

\author{Lakmali Nadeesha Kumari Rathukohe Mudiyanselage \and
        Sen-Ching Samson Cheung}
\institute{University of Kentucky, Lexington, KY 40506, USA\\
    \email{\{lakmali.nadeesha, sen-ching.cheung\}@uky.edu}}

\maketitle

\begin{abstract}
Frequency-based loss reweighting, the standard remedy for imbalanced
histopathology segmentation, implicitly assumes that rare classes are
difficult. Yet difficulty also arises from morphological variability, boundary
ambiguity, and contextual similarity, all largely orthogonal to class frequency. We propose \emph{Dynamic Focal Attention} (DFA), a simple, efficient
mechanism that learns class-specific difficulty directly within the
cross-attention of query-based mask decoders. DFA adds a learnable per-class
bias to the attention logits, reweighting representations \emph{before}
prediction rather than gradients \emph{after} it. Initialised from a centred
log-frequency prior to prevent gradient starvation and then optimised
end-to-end, the bias adapts to difficulty signals as they emerge during
training, unifying frequency- and difficulty-aware reweighting in a single
attention-bias framework. On three benchmarks (BCSS, BDSA, CRAG), DFA
consistently improves Dice and IoU, matching or exceeding a two-stage
difficulty-aware baseline without a separate estimator or extra training
stage. This shows that encoding difficulty at the representation level is a
principled alternative to conventional loss reweighting.

\keywords{Class difficulty \and Attention mechanism \and Pathology segmentation \and Transformer \and Difficulty-aware learning}
\end{abstract}

\section{Introduction}
\label{sec:intro}

Semantic segmentation of histopathology images underpins tissue
characterisation, tumour delineation, and biomarker extraction in computational
pathology~\cite{campanella2019clinical,lu2021data,graham2019hover}.
Transformer-based architectures~\cite{cheng2021per,cheng2022masked} perform
strongly overall yet underperform on clinically critical classes, a shortfall
typically attributed to class imbalance. But imbalance is only part of the story: class imbalance is a dataset-level frequency property, whereas class difficulty is governed by morphological variability, boundary ambiguity,
and contextual similarity.

\begin{figure*}[t]
  \centering
  \includegraphics[width=\textwidth]{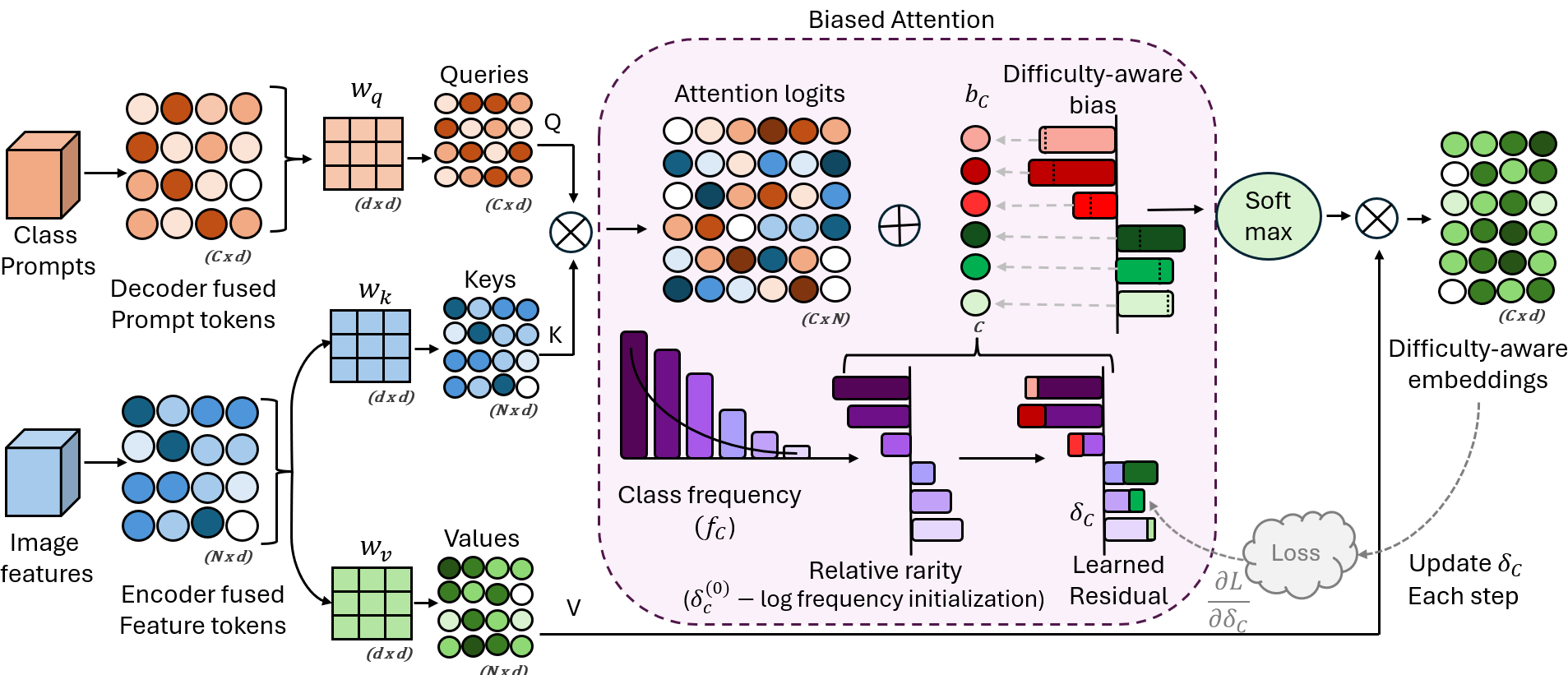}
  \caption{\textbf{DFA integrated into cross-attention.}
  A learnable class-specific bias $b_c$ is added to attention logits
  before softmax, initialised from a log-frequency prior and optimised
  end-to-end via $\mathcal{L}$ (Eq.~\ref{eq:loss}). Here
  $\otimes$ denotes matrix multiplication and $\oplus$ broadcast
  addition of $\mathbf{b}\!\in\!\mathbb{R}^{C}$ across $N$ spatial positions.}
  \label{fig:dfa_pipeline}
\end{figure*}

\noindent\textbf{Limitations of existing approaches.}\quad
Data-level strategies~\cite{buda2018systematic,kumari2025warmer} alter training
distributions without adding semantic information, while loss-level
methods~\cite{lin2017focal,yeung2022unified,cui2019class,shrivastava2016training}
reweight gradient magnitudes \emph{after} prediction, once class-specific
representations are already formed. All treat rarity as a proxy for difficulty,
yet the two are largely orthogonal~\cite{buda2018systematic,graham2023conic}
(\cref{sec:disconnect}). Complementary pathology methods target \emph{spatial}
difficulty, with HoVer-Net~\cite{graham2019hover} exploiting distance maps and
MILD-Net~\cite{graham2019mild} employing multi-scale dilated context, yet neither
supplies an explicit \emph{class-level} difficulty signal during feature
aggregation.

\noindent\textbf{Attention as a principled intervention point.}\quad
SAM~\cite{kirillov2023segment} and SAM-Path\allowbreak~\cite{zhang2023sam,chen2024uni}
recast segmentation as mask classification, with learnable class queries
interacting with image features through
cross-attention~\cite{carion2020end,cheng2022masked}---directly shaping
per-class feature aggregation and providing a natural site for difficulty-aware
intervention. Yet standard attention treats all queries uniformly. Structured
logit biases~\cite{raffel2020exploring,press2022train,zhu2021deformable} help
when task-specific priors are available, but none encode \emph{class-level
semantic difficulty}, which is crucial for the diverse challenges of
histopathology segmentation.

\noindent\textbf{Our approach.}\quad
We propose \textbf{Dynamic Focal Attention (DFA)}, which injects a learnable
class-specific scalar bias $b_c$ into cross-attention logits
(\cref{fig:dfa_pipeline}), enabling difficulty-aware feature aggregation at the
representation level. Rather than reweighting gradients after prediction, DFA
adjusts logits class-conditionally within attention, modulating how strongly
features are aggregated per class. To prevent early gradient starvation on rare
classes, $b_c$ is initialised from a centred log-frequency prior and thereafter
updated by the segmentation objective to emphasise difficult-to-segment classes.
We also formalise two variants, class-frequency (CFFA) and hard-class
(HCFA), giving a progression from static to dynamic difficulty estimation. DFA
runs in a single stage, adds one scalar per class, and needs no pre-trained
baseline.

\noindent\textbf{Contributions.}\quad
(\textit{1})~We show class frequency is an unreliable difficulty proxy, with
direct rarity--difficulty inversions across three pathology benchmarks.
(\textit{2})~We propose the \emph{Focal Attention Bias} framework, unifying
CFFA, HCFA, and DFA as a progressive family of class-difficulty-aware attention
mechanisms.
(\textit{3})~DFA achieves up to $15.2$ Dice points improvement on difficult classes and ${\sim}40\%$ training-time reduction versus two-stage approaches. The code is publicly available at \url{https://github.com/lakmali240/DFA-Imbalance}.

\section{Method}
\label{sec:method}

\subsection{Preliminaries: SAM-Path Decoder}
\label{sec:sampath}

We build on SAM-Path~\cite{zhang2023sam}, which extends
SAM~\cite{kirillov2023segment} with two encoders, a frozen SAM encoder and a pathology foundation model, whose outputs align into a unified feature
$\mathbf{H}\!\in\!\mathbb{R}^{N\times d}$ ($N$ spatial tokens, $d$ feature
dimension). The decoder keeps $C$ learnable class prompt tokens
$\mathcal{P}\!\in\!\mathbb{R}^{C \times d}$ that interact with $\mathbf{H}$ via
cross-attention to produce per-class masks $\hat{y}_c$ and IoU predictions
$\widehat{\mathrm{iou}}_c$, $c\!=\!1,\ldots,C$. The attention weights
$\alpha_{c,i}\!=\!\exp(s_{c,i})/\sum_{c'}\exp(s_{c',i})$ with
$s_{c,i}\!=\!(\mathbf{Q}\mathbf{K}^{\!\top})_{c,i}/\sqrt{d}$, where
$\mathbf{Q}\!\in\!\mathbb{R}^{C\times d}$ and
$\mathbf{K}\!\in\!\mathbb{R}^{N\times d}$ derive from $\mathcal{P}$ and
$\mathbf{H}$, depend only on feature--prompt similarity, with no mechanism to
modulate aggregation by class-level complexity, which is the limitation we address. Our formulation applies broadly to cross-attention in transformer-based
segmentation models beyond SAM-Path.

\subsection{Dynamic Focal Attention}
\label{sec:dfa}

\noindent\textbf{Focal Attention Bias framework.}\quad
We introduce a class-specific scalar $b_c\!\in\!\mathbb{R}$, broadcast across all
$N$ spatial positions and added to the cross-attention logit for class $c$ before a class-wise softmax at each spatial location:
\begin{equation}
  \tilde{\alpha}_{c,i}
    = \frac{\exp(s_{c,i}+b_c)}{\sum_{c'=1}^{C}\exp(s_{c',i}+b_{c'})},
  \quad
  \widetilde{\mathrm{Attn}}(\mathbf{Q},\mathbf{K},\mathbf{V})
    = \mathrm{softmax}\!\left(\tfrac{\mathbf{Q}\mathbf{K}^{\!\top}}
      {\sqrt{d}}\!\oplus\!\mathbf{b}\right)\mathbf{V},
  \label{eq:focal_attn}
\end{equation}
where $\mathbf{b}\!=\![b_1,\ldots,b_C]^\top\!\in\!\mathbb{R}^C$ and
$\mathbf{V}\!\in\!\mathbb{R}^{N\times d}$ are values from $\mathcal{P}$. The
operator $\oplus$ denotes broadcast addition of $\mathbf{b}$ across $N$ spatial
positions, following structured attention biases such as
ALiBi~\cite{press2022train}, T5~\cite{raffel2020exploring}, and Deformable
DETR~\cite{zhu2021deformable}. Setting
$b_c\!=\!\log\omega_c$ ($\omega_c\!>\!0$) multiplicatively rescales the $c$-th
score by $\omega_c$ uniformly across all aggregations; parameterising $b_c$
yields a progressive family of strategies.

\noindent\textbf{(S1)~Class Frequency Focal Attention (CFFA).}\quad
CFFA sets $b_c$ inversely proportional to pixel frequency:
\begin{equation}
  b_c = \gamma\,\log(1-f_c), \qquad \gamma>0,
  \label{eq:cffa}
\end{equation}
where $f_c\!\in\!(0,1)$ is the normalised training-set pixel frequency of class
$c$ and $\gamma$ controls magnitude. Unlike output-level logit
adjustment~\cite{menon2021logit}, this prior acts on feature aggregation
directly, but still conflates rarity with difficulty (\cref{sec:disconnect}).

\noindent\textbf{(S2)~Hard Class Focal Attention (HCFA).}\quad
HCFA replaces frequency with empirical difficulty estimated from baseline performance:
\begin{equation}
  b_c = \gamma\,\log(1-p_c),
  \quad
  p_c = \frac{\mathrm{Dice}_c^{\mathrm{base}}}
             {\sum_{c'=1}^{C}\mathrm{Dice}_{c'}^{\mathrm{base}}},
  \label{eq:hcfa}
\end{equation}
where $\mathrm{Dice}_c^{\mathrm{base}}$ is the per-class validation Dice of the
pre-trained baseline and $p_c$ its normalised difficulty proxy; HCFA targets
hard classes but requires a full pre-training run.

\noindent\textbf{(S3)~Dynamic Focal Attention (DFA)---Proposed.}\quad
CFFA and HCFA fix $b_c$ before training and cannot adapt to difficulty signals
that emerge during optimisation. DFA instead treats $b_c$ as a learnable scalar
$\delta_c\!\in\!\mathbb{R}$ optimised end-to-end:
\begin{equation}
  b_c = \delta_c, \qquad \boldsymbol{\delta}\in\mathbb{R}^C,
  \label{eq:dfa}
\end{equation}
\begin{equation}
  \left.\frac{\partial\tilde{\alpha}_{c,i}}{\partial\delta_c}\right|_{\text{on-diag.}}
    = \underbrace{\tilde{\alpha}_{c,i}(1-\tilde{\alpha}_{c,i})}_{\text{self-gating factor}},
  \quad
  \frac{\partial\mathcal{L}}{\partial\delta_c}
    = \sum_{k=1}^{C}\sum_{i=1}^{N}
      \frac{\partial\mathcal{L}}{\partial\tilde{\alpha}_{k,i}}\,
      \tilde{\alpha}_{k,i}\bigl(\mathbf{1}[k\!=\!c]-\tilde{\alpha}_{c,i}\bigr),
  \label{eq:grad}
\end{equation}
driving $\delta_c$ positive for poorly-segmented classes and suppressing it for
well-segmented ones. The on-diagonal self-gating factor
$\tilde{\alpha}_{c,i}(1\!-\!\tilde{\alpha}_{c,i})$ peaks when attention weights
are uncertain and vanishes at saturation, yielding an implicit curriculum effect.

\noindent\textbf{Warm-start initialisation.}\quad
Zero-initialising $\boldsymbol{\delta}$ starves rare classes of gradient
($\tilde{\alpha}_{c,i}\!\approx\!0\Rightarrow$ self-gating $\approx\!0$,
regardless of loss). We therefore initialise $\boldsymbol{\delta}_c$ from a
\emph{centred log-frequency prior}:
\begin{equation}
  \delta_c^{(0)} = \beta\!\left(
    \log\pi_c - \tfrac{1}{C}\sum_{c'=1}^{C}\log\pi_{c'}
  \right),
  \quad \log\pi_c = \gamma\,\log(1-f_c),
  \label{eq:warmstart}
\end{equation}
giving minority classes a positive initial bias and majority
classes a negative one at zero mean shift (in practice $\gamma\!=\!2$,
$\beta\!=\!0.8$). Thereafter $\delta_c$ is learned end-to-end with no explicit
frequency or hardness constraints.

\noindent\textbf{Training objective.}\quad
Following SAM-Path~\cite{zhang2023sam}, the objective sums per-class soft Dice
loss $\mathcal{L}_{\mathrm{dice}}$, focal loss~\cite{lin2017focal}
$\mathcal{L}_{\mathrm{focal}}$, and MSE-IoU regression
$\mathcal{L}_{\mathrm{mse}}$ against the realised
$\mathrm{IoU}(\hat{y}_c,y_c)\!=\!|\hat{y}_c\cap y_c|/|\hat{y}_c\cup y_c|$, plus an
$\ell_2$ regulariser on the learnable biases:
\begin{equation}
  \mathcal{L}
    = \sum_{c=1}^{C}\!\bigl[
        (1-\alpha)\mathcal{L}_{\mathrm{dice}}(\hat{y}_c,y_c)
       +\alpha\,\mathcal{L}_{\mathrm{focal}}(\hat{y}_c,y_c)
       +\mu\,\mathcal{L}_{\mathrm{mse}}\!\bigl(\widehat{\mathrm{iou}}_c,
         \mathrm{IoU}(\hat{y}_c,y_c)\bigr)
      \bigr]
      + \lambda\|\boldsymbol{\delta}\|_2^2,
  \label{eq:loss}
\end{equation}
where $\alpha\!\in\![0,1]$ balances Dice and focal terms, $\mu\!>\!0$ weights IoU
regression, and $\lambda\!=\!10^{-4}$ prevents unbounded bias growth. We use a
$100\times$ higher learning rate $\eta_b$ for $\delta_c$ for rapid adaptation.
DFA adds negligible cost: one scalar per class and one broadcast logit shift per
cross-attention layer.

\section{Experiments and Results}
\label{sec:experiments}

\subsection{Experimental Setup}
\label{sec:setup}

\noindent\textbf{Datasets and splits.}\quad
We evaluate on three public histopathology benchmarks, each using the official
train/test split with $20\%$ of training held out for validation
(SAM-Path protocol~\cite{zhang2023sam}); a single fixed split (no
cross-validation) is held identical across all methods.
(\textbf{i})~\textbf{BCSS}~\cite{amgad2019structured}: breast cancer, five
classes (Tumor, Stroma, Inflammatory, Necrosis, Other), $40\!\times$,
$1024^2$.
(\textbf{ii})~\textbf{BDSA}~\cite{flanagan2025bdsa}: brain tissue, five classes
(Background, Gray/White Matter, Leptomeninges, Superficial Cortex),
$10\!\times$, $1024^2$; $10$ WSIs yield ${\sim}82\text{k}$ patches, split at the
slide level.
(\textbf{iii})~\textbf{CRAG}~\cite{graham2019mild}: colorectal adenocarcinoma,
two classes (Non-gland, Gland), $20\!\times$, $1536^2$.

\noindent\textbf{Architecture and training.}\quad
All models use SAM-Path with ViT-B (SAM encoder) and HIPT (pathology encoder),
trained with AdamW ($\eta\!=\!10^{-4}$, batch~$4$, cosine annealing) on NVIDIA
A100 GPUs. Epoch counts ($30$\allowbreak/$20$\allowbreak/$60$ for BCSS\allowbreak/BDSA\allowbreak/CRAG) follow each
dataset's validation-loss plateau and are identical across all five methods, so
any residual schedule effect is shared.
\textbf{Hyperparameters:} $\gamma\!=\!2$ for all focal variants;
$\alpha\!=\!0.25$, $\mu\!=\!0.0625$ on BCSS/BDSA and $\alpha\!=\!0.125$,
$\mu\!=\!0$ on CRAG (SAM-Path defaults); $\beta\!=\!0.8$, $\lambda\!=\!10^{-4}$,
$\eta_b\!=\!10^{-2}$ for DFA, all fixed across datasets after a small grid
search on BCSS, with stable performance and no extensive tuning.

\noindent\textbf{Baselines.}\quad
Five configurations differing \emph{only} in attention bias:
(\textbf{i})~\emph{Base w/o FL} (Eq.~\ref{eq:loss} without focal term);
(\textbf{ii})~\emph{Base w/ FL} (SAM-Path default);
(\textbf{iii})~\emph{CFFA} (Eq.~\ref{eq:cffa});
(\textbf{iv})~\emph{HCFA} (two-stage, Eq.~\ref{eq:hcfa});
(\textbf{v})~\emph{DFA (Ours)} (single-stage, Eq.~\ref{eq:dfa}).

\subsection{The Frequency--Difficulty Disconnect}
\label{sec:disconnect}
\begin{figure}[t]
  \centering
  \includegraphics[width=\linewidth]{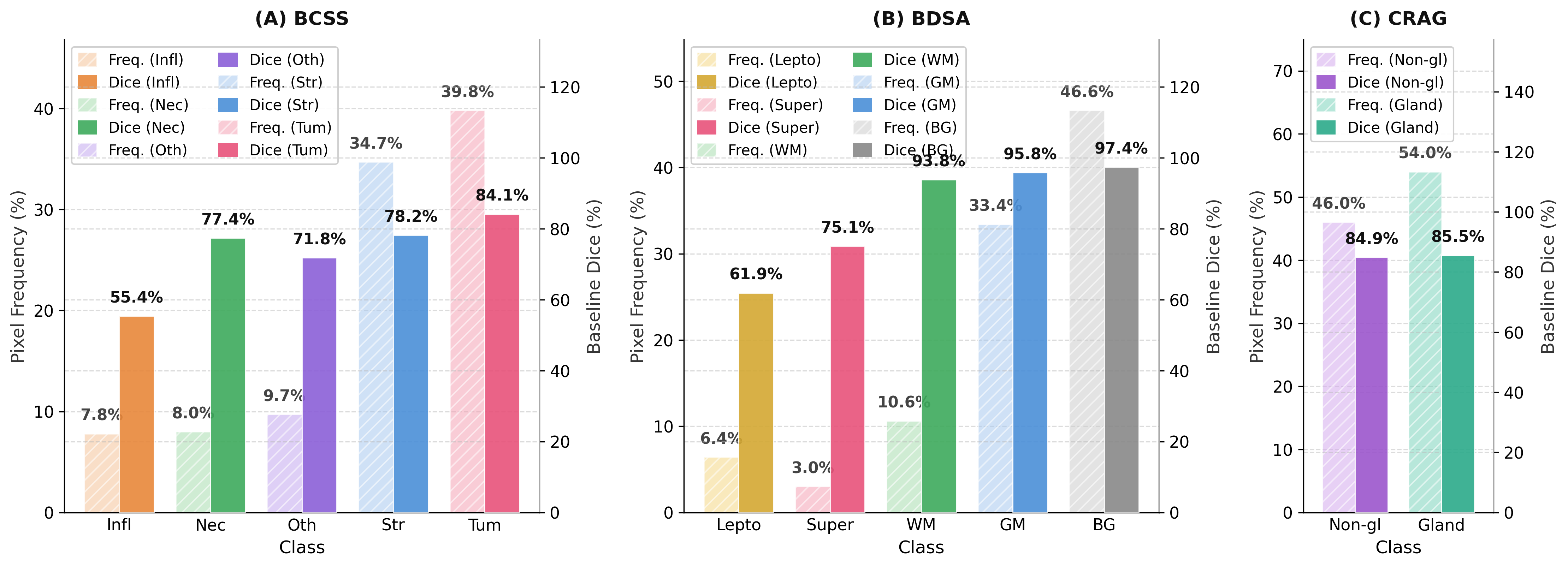}
  \caption{\textbf{Frequency--difficulty disconnect.}
  Pixel frequency (hatched bars) vs.\ Dice of the focal-loss baseline
  (solid bars) per class across three datasets.}
  \label{fig:freq_disconnect}
\end{figure}

We first test whether pixel frequency predicts segmentation difficulty---the
assumption behind CFFA and frequency-based reweighting. The focal-loss baseline
contradicts it across all three benchmarks (\cref{fig:freq_disconnect}). In
BCSS, Inflammatory and Necrosis share near-identical frequency yet differ by
$22.0$ Dice points; Other (the most frequent minority class) scores below the
rarer Necrosis ($71.8\%$ vs.\ $77.4\%$), inverting the rarity--difficulty
ranking. In BDSA, the rarest class (Superficial Cortex, $3.0\%$) beats the more
frequent Leptomeninges ($6.4\%$) by $13.2$ points. Only CRAG, near-balanced,
shows frequency loosely tracking difficulty.

Thus true difficulty is governed by morphological variability, boundary
ambiguity, and inter-class contextual confusion~\cite{buda2018systematic},
largely independent of pixel frequency. This is the regime in which we evaluate
CFFA, HCFA, and DFA.

\subsection{Main Results}
\label{sec:results}
\cref{tab:all_results} reports per-class Dice (F1 overlap) and IoU (Jaccard);
IoU mirrors Dice throughout, so discussion focuses on Dice, with hard classes
(baseline Dice below the dataset mean) of primary interest. Since CFFA already
captures the frequency-related component of difficulty, the CFFA$\rightarrow$DFA
gain concentrates on the non-frequency residual (i.e., hard classes) rather
than on global mean Dice.

\noindent\textbf{BCSS.}\quad
DFA reaches the highest mean Dice ($0.7826$), $+4.9$ over the focal-loss baseline
and $+0.5$ over HCFA. Inflammatory improves $+15.2$ over Base w/o FL and $+2.7$
over CFFA, confirming DFA targets true difficulty rather than frequency; it also
gives the best Other ($0.7581$), whose difficulty frequency-based methods
underestimate.

\noindent\textbf{BDSA.}\quad
Without focal loss the base model collapses on Leptomeninges and Superficial
Cortex (Dice $=\!0.000$); focal loss recovers them to $0.6191$ and $0.7508$. DFA
further raises both to $0.6908$ and $0.7808$ and achieves the best White Matter
($0.9556$).

\noindent\textbf{CRAG.}\quad
Even with minimal imbalance, DFA attains the highest mean Dice ($0.8858$),
$+0.5$ over HCFA and $+1.5$ over CFFA, capturing difficulty structure beyond
frequency.

\noindent\textbf{Efficiency.}\quad
As a single-stage method, DFA cuts wall-clock training time by ${\sim}40\%$ versus two-stage HCFA while matching or exceeding its Dice everywhere.

\begin{table*}[t]
  \centering
  \caption{%
    \textbf{Segmentation performance across three histopathology benchmarks.}
    Each method occupies two columns: Dice (left) and IoU (right). Means and stds are computed over three independent training runs with different random seeds. Hard classes (baseline Dice below dataset mean) are shaded
    \colorbox[HTML]{FFF3CD}{amber}.
    \textbf{Bold}: best result per row.
  }
  \label{tab:all_results}
  \resizebox{\textwidth}{!}{%
  \begin{tabular}{ll|cc|cc|cc|cc|cc}
    \toprule
    & &
    \multicolumn{2}{c|}{\textbf{Base w/o FL}~\cite{zhang2023sam}} &
    \multicolumn{2}{c|}{\textbf{Base w/ FL}~\cite{zhang2023sam}} &
    \multicolumn{2}{c|}{\textbf{CFFA}\textsuperscript{\cite{menon2021logit}}} &
    \multicolumn{2}{c|}{\textbf{HCFA}(Ours)} &
    \multicolumn{2}{c}{\textbf{DFA (Ours)}} \\
    \cmidrule(lr){3-4}\cmidrule(lr){5-6}\cmidrule(lr){7-8}
    \cmidrule(lr){9-10}\cmidrule(lr){11-12}
    \textbf{Dataset} & \textbf{Class}
      & Dice & IoU & Dice & IoU & Dice & IoU & Dice & IoU & Dice & IoU \\
    \midrule
    \multirow{7}{*}{\textbf{BCSS}}
      & \textit{Mean}
        & 0.7032 & 0.5535 & 0.7336 & 0.5882 & 0.7737 & 0.6382
        & 0.7780 & 0.6434 & \best{0.7826} & \best{0.6488} \\
      & \small$\pm$\textit{std}
        & \small$\pm 0.003$ & \small$\pm 0.004$ & \small$\pm 0.003$ & \small$\pm 0.003$
        & \small$\pm 0.002$ & \small$\pm 0.003$ & \small$\pm 0.002$ & \small$\pm 0.003$
        & \small$\pm 0.002$ & \small$\pm 0.002$ \\
      & Tumor
        & 0.8298 & 0.7090 & 0.8410 & 0.7257 & 0.8662 & 0.7640
        & 0.8674 & 0.7659 & \best{0.8684} & \best{0.7674} \\
      & Stroma
        & 0.7521 & 0.6027 & 0.7816 & 0.6415 & 0.7968 & 0.6622
        & 0.7944 & 0.6590 & \best{0.7969} & \best{0.6624} \\
      & \cellcolor{hardcol}Inflamm.\!\!
        & \cellcolor{hardcol}0.4974 & \cellcolor{hardcol}0.3310
        & \cellcolor{hardcol}0.5537 & \cellcolor{hardcol}0.3828
        & \cellcolor{hardcol}0.6228 & \cellcolor{hardcol}0.4522
        & \cellcolor{hardcol}0.6392 & \cellcolor{hardcol}0.4697
        & \cellcolor{hardcol}\best{0.6493} & \cellcolor{hardcol}\best{0.4807} \\
      & Necrosis
        & 0.7615 & 0.6148 & 0.7739 & 0.6312 & 0.8311 & 0.7110
        & 0.8398 & \best{0.7238} & \best{0.8402} & 0.7230 \\
      & \cellcolor{hardcol}Other
        & \cellcolor{hardcol}0.6754 & \cellcolor{hardcol}0.5099
        & \cellcolor{hardcol}0.7178 & \cellcolor{hardcol}0.5598
        & \cellcolor{hardcol}0.7514 & \cellcolor{hardcol}0.6018
        & \cellcolor{hardcol}0.7490 & \cellcolor{hardcol}0.5987
        & \cellcolor{hardcol}\best{0.7581} & \cellcolor{hardcol}\best{0.6104} \\
    \midrule
    \multirow{7}{*}{\textbf{BDSA}}
      & \textit{Mean}
        & 0.5589 & 0.5241 & 0.8480 & 0.7644 & 0.8703 & 0.7931
        & 0.8712 & 0.7944 & \best{0.8739} & \best{0.7953} \\
      & \small$\pm$\textit{std}
        & \small$\pm 0.003$ & \small$\pm 0.003$ & \small$\pm 0.002$ & \small$\pm 0.003$
        & \small$\pm 0.002$ & \small$\pm 0.002$ & \small$\pm 0.002$ & \small$\pm 0.002$
        & \small$\pm 0.001$ & \small$\pm 0.002$ \\
      & BG
        & 0.9530 & 0.9121 & 0.9742 & 0.9501 & 0.9761 & 0.9537
        & \best{0.9763} & \best{0.9539} & \best{0.9763} & 0.9538 \\
      & GM
        & 0.9270 & 0.8644 & 0.9585 & 0.9205 & 0.9648 & 0.9322
        & 0.9656 & \best{0.9337} & \best{0.9658} & 0.9335 \\
      & WM
        & 0.9147 & 0.8441 & 0.9375 & 0.8834 & 0.9504 & 0.9060
        & 0.9514 & 0.9079 & \best{0.9556} & \best{0.9090} \\
      & \cellcolor{hardcol}Lepto.
        & \cellcolor{hardcol}0.0000 & \cellcolor{hardcol}0.0000
        & \cellcolor{hardcol}0.6191 & \cellcolor{hardcol}0.4660
        & \cellcolor{hardcol}0.6840 & \cellcolor{hardcol}0.5384
        & \cellcolor{hardcol}0.6851 & \cellcolor{hardcol}0.5395
        & \cellcolor{hardcol}\best{0.6908} & \cellcolor{hardcol}\best{0.5405} \\
      & \cellcolor{hardcol}Super.
        & \cellcolor{hardcol}0.0000 & \cellcolor{hardcol}0.0000
        & \cellcolor{hardcol}0.7508 & \cellcolor{hardcol}0.6022
        & \cellcolor{hardcol}0.7761 & \cellcolor{hardcol}0.6351
        & \cellcolor{hardcol}0.7773 & \cellcolor{hardcol}0.6368
        & \cellcolor{hardcol}\best{0.7808} & \cellcolor{hardcol}\best{0.6395} \\
    \midrule
    \multirow{4}{*}{\textbf{CRAG}}
      & \textit{Mean}
        & 0.8423 & 0.7276 & 0.8520 & 0.7422 & 0.8707 & 0.7711
        & 0.8806 & 0.7868 & \best{0.8858} & \best{0.7951} \\
      & \small$\pm$\textit{std}
        & \small$\pm 0.003$ & \small$\pm 0.004$ & \small$\pm 0.002$ & \small$\pm 0.003$
        & \small$\pm 0.002$ & \small$\pm 0.003$ & \small$\pm 0.002$ & \small$\pm 0.002$
        & \small$\pm 0.002$ & \small$\pm 0.002$ \\
      & \cellcolor{hardcol}Non-gland
        & \cellcolor{hardcol}0.8392 & \cellcolor{hardcol}0.7229
        & \cellcolor{hardcol}0.8487 & \cellcolor{hardcol}0.7372
        & \cellcolor{hardcol}0.8643 & \cellcolor{hardcol}0.7611
        & \cellcolor{hardcol}0.8712 & \cellcolor{hardcol}0.7717
        & \cellcolor{hardcol}\best{0.8774} & \cellcolor{hardcol}\best{0.7816} \\
      & Gland
        & 0.8454 & 0.7322 & 0.8553 & 0.7472 & 0.8771 & 0.7812
        & 0.8901 & 0.8020 & \best{0.8942} & \best{0.8087} \\
    \bottomrule
  \end{tabular}
  }
\end{table*}

\begin{figure}[t]
  \centering
  \includegraphics[width=\linewidth]{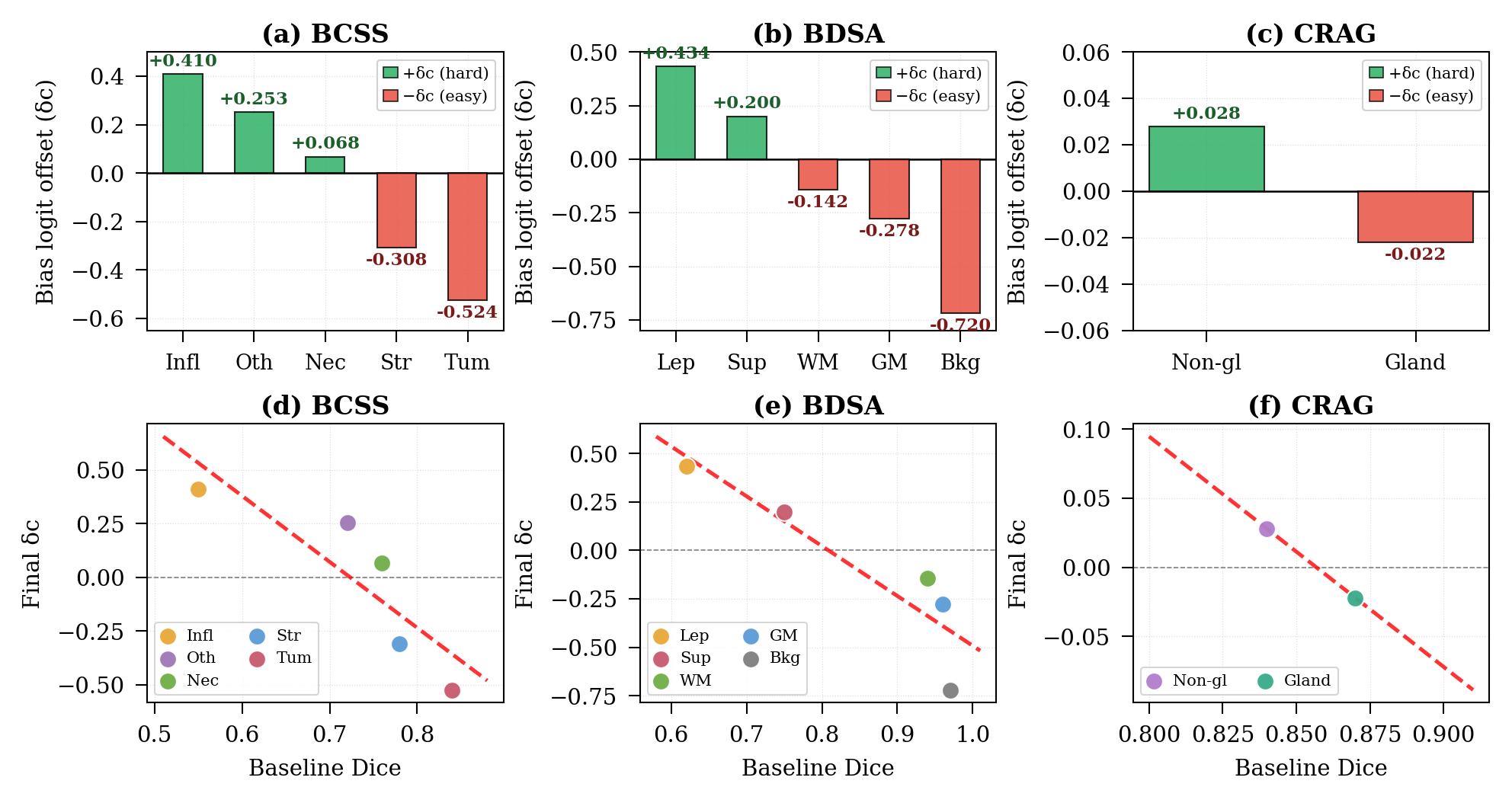}
  \caption{\textbf{Converged biases $\delta_c$ and difficulty correlation.}
  Top (a--c): $\delta_c$ per class
  (\textcolor{green!60!black}{\textbf{positive}} $\Leftrightarrow$ hard,
  \textcolor{red!70!black}{\textbf{negative}} $\Leftrightarrow$ easy).
  Bottom (d--f): $\delta_c$ vs.\ baseline Dice with fitted trend line;
  the strong negative correlation indicates $\delta_c$ tracks empirical
  difficulty rather than frequency.}
  \label{fig:bias_analysis}
\end{figure}

\begin{table}[t]
\centering
\caption{Warm- vs.\ cold-start ablation on BCSS.}
\label{tab:warmstart}
\setlength{\tabcolsep}{3pt}
\renewcommand{\arraystretch}{0.75}
\resizebox{\linewidth}{!}{%
\scriptsize
\begin{tabular}{l|rrrrr|r}
\toprule
\textbf{Init} & $\delta_\text{Infl.}$ & $\delta_\text{Oth.}$ & $\delta_\text{Nec.}$ & $\delta_\text{Str.}$ & $\delta_\text{Tum.}$ & \textbf{Dice\;($\pm$std)} \\
\midrule
Cold ($\delta_c^{(0)}\!=\!0$) & $+0.010$ & $+0.002$ & $+0.001$ & $-0.012$ & $-0.019$ & $0.7751_{\pm.005}$ \\
Warm (Eq.~\ref{eq:warmstart}) & $+0.410$ & $+0.253$ & $+0.068$ & $-0.308$ & $-0.524$ & $0.7826_{\pm.002}$ \\
\bottomrule
\end{tabular}}
\end{table}

\subsection{Analysis of Learned Attention Biases}
\label{sec:bias_analysis}
At convergence (\cref{fig:bias_analysis}a--c), $\boldsymbol{\delta}$ provides a
model-derived difficulty signal: hard classes acquire positive bias (e.g.,
Inflammatory $+0.410$; Leptomeninges $+0.434$), easy classes negative (e.g.,
Tumor $-0.524$; Background $-0.720$). The scatter plots
(\cref{fig:bias_analysis}d--f), plotting $\delta_c$ against \emph{baseline
Dice} rather than frequency, show a strong negative correlation between
difficulty and $\delta_c$, complementary to, not contradicting, the
frequency--difficulty disconnect in \cref{fig:freq_disconnect}. Crucially,
biases diverge from their frequency-based init: Inflammatory ($f_c\!=\!7.8\%$)
gets $+0.410$ vs.\ Necrosis $+0.068$ at near-identical frequency ($8.0\%$), a
$6\times$ gap mirroring their $22$-point Dice difference, which no
frequency-driven scheme could produce. Thus, $\delta_c$ captures difficulty sources, such as boundary complexity and contextual confusion, that are invisible to frequency alone.

\noindent\textbf{Ablation: warm-start initialisation.}\quad
\cref{tab:warmstart} isolates the centred log-frequency prior
(Eq.~\ref{eq:warmstart}) on BCSS. Cold-start converges to a near-flat $\delta$
profile ($|\delta_c|\!\leq\!0.02$) as self-gating suppression
($\tilde{\alpha}_{c,i}(1-\tilde{\alpha}_{c,i})\!\approx\!0$) stalls updates.
Warm-start breaks this degeneracy: ${\sim}20{\times}$ wider spread and $+0.75$
Dice. Beyond the Dice gain, the warm prior is what lets DFA learn the intended
difficulty-aware profile (hard $+$, easy $-$); without it $\delta$ stays
uninformative even when overall Dice is close.

\subsection{Qualitative Results}
\label{sec:qualitative}
\begin{figure*}[t]
  \centering
  \setlength{\tabcolsep}{1.5pt}
  \renewcommand{\arraystretch}{0.8}
  \resizebox{\textwidth}{!}{%
  \begin{tabular}{>{\centering\arraybackslash}m{1.5em}ccccccc}
    &
    \scriptsize Image (WSI) &
    \scriptsize Ground Truth &
    \scriptsize w/o FL &
    \scriptsize w/ FL &
    \scriptsize CFFA &
    \scriptsize HCFA &
    \scriptsize\textbf{DFA (Ours)} \\[1pt]
    \rotatebox[origin=c]{90}{\parbox{0.128\linewidth}{\centering\footnotesize\textbf{BCSS}}} &
    \includegraphics[width=0.128\linewidth]{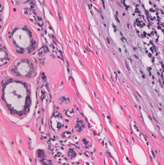} &
    \includegraphics[width=0.128\linewidth]{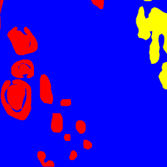} &
    \includegraphics[width=0.128\linewidth]{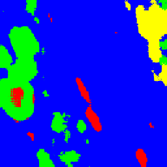} &
    \includegraphics[width=0.128\linewidth]{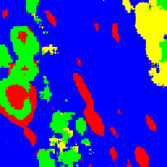} &
    \includegraphics[width=0.128\linewidth]{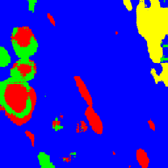} &
    \includegraphics[width=0.128\linewidth]{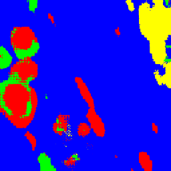} &
    \includegraphics[width=0.128\linewidth]{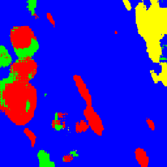} \\[2pt]
    \rotatebox[origin=c]{90}{\parbox{0.128\linewidth}{\centering\footnotesize\textbf{CRAG}}} &
    \includegraphics[width=0.128\linewidth]{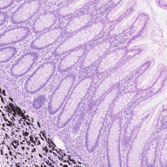} &
    \includegraphics[width=0.128\linewidth]{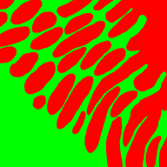} &
    \includegraphics[width=0.128\linewidth]{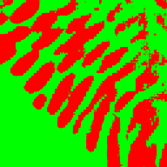} &
    \includegraphics[width=0.128\linewidth]{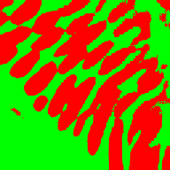} &
    \includegraphics[width=0.128\linewidth]{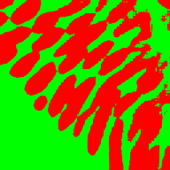} &
    \includegraphics[width=0.128\linewidth]{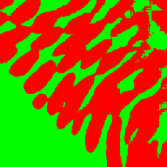} &
    \includegraphics[width=0.128\linewidth]{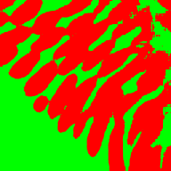} \\[2pt]
    \rotatebox[origin=c]{90}{\parbox{0.128\linewidth}{\centering\footnotesize\textbf{BDSA}}} &
    \includegraphics[width=0.128\linewidth]{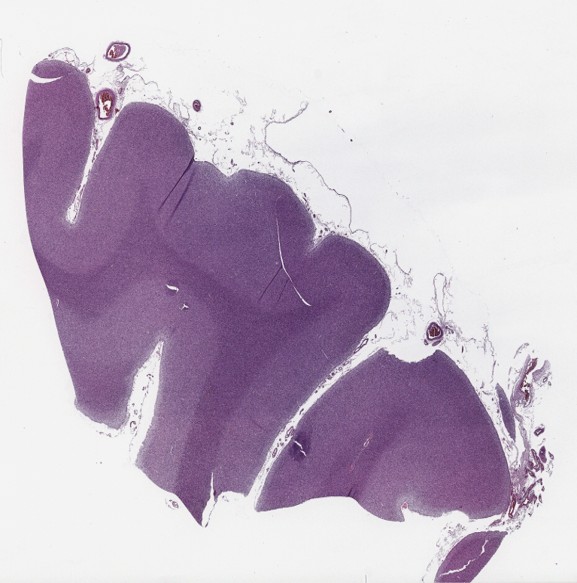} &
    \includegraphics[width=0.128\linewidth]{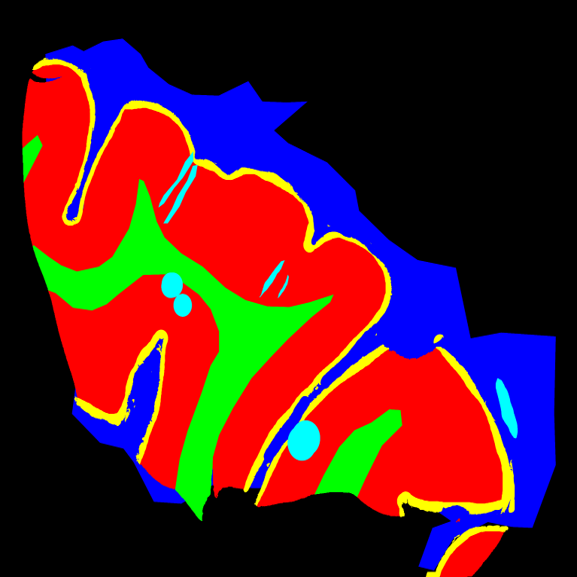} &
    \includegraphics[width=0.128\linewidth]{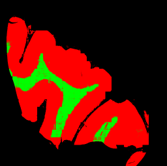} &
    \includegraphics[width=0.128\linewidth]{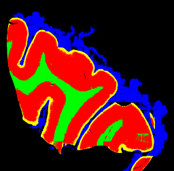} &
    \includegraphics[width=0.128\linewidth]{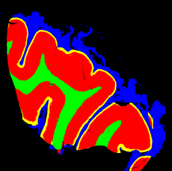} &
    \includegraphics[width=0.128\linewidth]{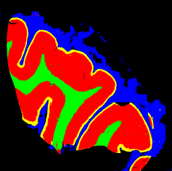} &
    \includegraphics[width=0.128\linewidth]{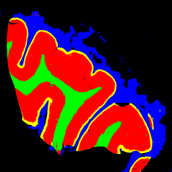} \\
  \end{tabular}
  }
  \caption{%
    \textbf{Qualitative segmentation comparison across three pathology
    benchmarks.}
    Each row shows a WSI patch with ground truth and predictions from
    all five methods.\\[4pt]
    {\footnotesize\textbf{BCSS:}}\;\swatch{FF0000}{Inflamm.}\;\swatch{00AA00}{Tumor}\;\swatch{0000FF}{Stroma}\;\swatch{FFFF00}{Other}\;\swatch{FF00FF}{Necrosis}\quad
    {\footnotesize\textbf{CRAG:}}\;\swatch{00CC00}{Non-Gland}\;\swatch{FF0000}{Gland}\quad
    {\footnotesize\textbf{BDSA:}}\;\swatch{000000}{BG}\;\swatch{FF0000}{Gray M.}\;\swatch{00CC00}{White M.}\;\swatch{0000FF}{Leptom.}\;\swatch{FFFF00}{Superficial}.
  }
  \label{fig:qualitative}
\end{figure*}

\cref{fig:qualitative} shows consistent gains: misclassified Inflammatory
regions in BCSS are recovered with coherent boundaries; Leptomeninges boundaries
in BDSA sharpen progressively from CFFA to HCFA to DFA; and CRAG glands show
markedly reduced fragmentation.

\section{Limitations and Future Work}
DFA applies a \emph{single global} bias per class. Classes such as Tumor or
Stroma can be harder in some regions than others, so a class-level scalar
captures only \emph{average} difficulty, acting as a prior, while the
token-dependent $\mathbf{Q}\mathbf{K}^{\!\top}$ term preserves spatial
selectivity. Spatially adaptive or region-conditioned biases capturing
instance-level and intra-class difficulty are a natural extension. Broader
validation across architectures (CNN--transformer hybrids, other query-based
decoders) and modalities (MRI, CT), explicit attention-map comparison, and a
convergence analysis remain future work.

\section{Conclusion}
\label{sec:conclusion}
Focal loss alone is insufficient and frequency-based reweighting unreliable:
CFFA fails when rare classes are not inherently hard, and HCFA needs a costly
pre-training stage. DFA resolves both by learning class-specific difficulty
continuously within cross-attention, outperforming all baselines across three
benchmarks with only $C$ extra scalar parameters. The warm-start prior prevents
gradient starvation, the self-gating factor gives implicit curriculum
regulation, and encoding difficulty at the representation level yields
model-wide gains over post-hoc gradient corrections.

\begin{credits}
\subsubsection{\ackname}
The work was supported by NIH grants P30 AG072946 and U24 NS133945. We are
grateful to the research participants, clinicians, and staff for making this
work possible. This work used Delta advanced computing and data resource at the
University of Illinois Urbana-Champaign through allocation CIS230383 from the
Advanced Cyberinfrastructure Coordination Ecosystem: Services \& Support
(ACCESS) program, which is supported by U.S. National Science Foundation grants
\#2138259, \#2138286, \#2138307, \#2137603, and \#2138296.

\subsubsection{\discintname}
The authors have no competing interests to declare that are relevant to the
content of this article.
\end{credits}

\bibliographystyle{splncs04}
\bibliography{Paper-6050}

\end{document}